# FEASIBILITY STUDY OF CHANNELING ACCELERATION EXPERIMENT AT THE FERMILAB ASTA FACILITY

[1, 3]Young-Min Shin, [2]Tao Xu, [3]Dean A. Still, and [3]Vladimir Shiltsev
[1]Department of Physics, Northern Illinois University, Dekalb, IL, 60115, USA
[2]Department of Chemistry, Northern Illinois University, Dekalb, IL, 60115, USA
[3]Fermi National Accelerator Laboratory (FNAL), Batavia, IL 60510, USA

*Abstract*

Crystal channeling technology has offered various opportunities in accelerator community with a viability of ultrahigh gradient (TV/m) acceleration for future HEP collider in Energy Frontier. The major challenge of the channeling acceleration is that ultimate acceleration gradients might require high power driver at hard x-ray regime (~ 40 keV), exceeding those conceivable for x-rays as of today, though x-ray lasers can efficiently excite solid plasma and accelerate particles inside a crystal channel. Moreover, only disposable crystal accelerators are possible at such high externally excited fields which would exceed the ionization thresholds destroying the atomic structure, so acceleration will take place only in a short time before full dissociation of the lattice. Carbon-based nanostructures have great potential with a wide range of flexibility and superior physical strength, which can be applied to channeling acceleration. This paper present beam-driven channeling acceleration concept with CNTs and discuss feasible experiments with the Advanced Superconducting Test Area (ASTA) in Fermilab and beyond.

## INTRODUCTION

The cost models of the modern colliders are quite complicated, but one may safely assume that a future facility should not exceed a few tens of km in length and simultaneously require less than 10 to a few tens of MW of beam. To get to the energies of interest within the given footprint, fast particle acceleration is inevitable. Plasma-wakefield acceleration (PWA) has become of great interest because of the promise to offer extremely large acceleration gradients, on the order of $E_0 \approx n_0^{1/2}$ [GeV/m], where $n_0$ is the ambient electron number density ($n_0$ [$10^{18}$cm$^{-3}$]), on the order of 30-100 GV/m at plasma densities of $n_0 = 10^{17} – 10^{18}$ cm$^{-3}$.[1] The density of charge carriers (conduction electrons) in solids $n_0 = \sim 10^{20} – 10^{23}$ cm$^{-3}$ is significantly higher than what was considered above in plasma, and correspondingly, wakefields of up to 100 GeV/cm or 10 TV/m are possible. In the solid plasma, as escaping from a driving field due to fast pitch-angle diffusion resulting from increased scattering rates, particles must be accelerated along major crystallographic directions. This is called "channeling acceleration". Normally, crystal channeling has been applied to high energy beam control such as collimation, bending, and refraction [2]. For high energy beam optics, carbon nanotubes (CNTs) have been considered for bending and collimation [3, 4] on account of the much wider range of flexibility, including superior physical strength, which also ideally fits with channeling acceleration. CNTs, composed of graphene sheets rolled into seamless hollow cylinders with diameters ranging from 1nm to about sub-micron, exhibit unique physical and chemical properties as a quasi-one dimensional material [5]. In principle, both straight and bent CNTs can effectively be used for high-energy particle channeling [6], provided the technological challenge of achieving an almost perfect alignment of CNTs with respect to the beam direction could be tackled effectively in the synthesis of samples.

Recently, Fermilab built the Advanced Superconducting Test Accelerator (ASTA) facility (50 MeV and several hundreds of MeV energy beams) that will enable a broad range of electron beam-based experiments to study fundamental limitations to beam intensity and to developing transformative approaches to particle-beam generation, acceleration and manipulation, which is ideally suited for the channeling acceleration experiment. We plan to detect a measurable energy gain from the electron bunches passing through CNTs. Successful demonstration of the experiment with this beam driven method will verify the viability of CNT channeling interaction for ultra-high gradient acceleration, which will also prove the feasibility of the laser-driven channeling acceleration. The experimental setup will be accommodated to a 50 MeV main stream beamline and high energy (50 – 300 MeV) beamline in our plan. In the research, beam energies and radiation spectra of CNT samples will be mainly characterized by beam tests at relativistic regimes.

## CHANNELING ACCELERATION

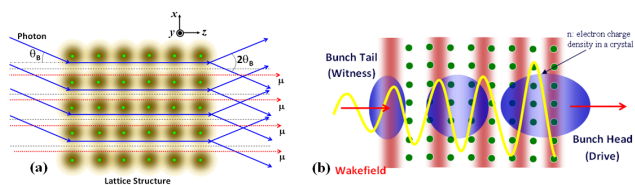

Figure 1: (a) Laser (x-ray) driven acceleration (muon, proton) (b) beam-driven acceleration (electron).

Wakefields in crystals can be excited by two sorts of driving sources: x-ray laser (Fig. 1(a)) or short electron bunch (Fig. 1(b)). With the x-ray pumping method [7], a crystal channel can hold > $10^{13}$ V/cm transverse and $10^9$ V/cm longitudinal fields of diffracted traveling EM-waves at the Bragg diffraction condition ($\lambda/2b = \sin\theta_B$, where b is the lattice constant and $\theta_B$ is the diffraction angle) However, to hold the ultimate gradients, the acceleration requires coherent hard x-rays ($\hbar\omega \approx 40$ keV) of $\geq 3$ GW to compensate for radiation losses, which exceed those conceivable today. The x-ray driving method thus fits for heavy particles, e.g. muons and protons, which have relatively smaller radiation losses. For electrons, the beam-driven acceleration is more favorably applicable to channeling acceleration as the energy losses of a drive beam can be transformed into acceleration energy of a witness beam [8]. The highly intensive plasma interaction in a crystal channel induces thermal radiations and collisional impacts accompanied by a large amount of heat energy, which would exceed the ionization thresholds and may even destroy the atomic structure. Only disposable forms of crystals such as fibers or films can be used for channeling acceleration. Also, lattice structures of crystals have fixed atomic dimensions, which thereby have some limits in designing acceleration parameters to mitigate physical constraints in solid plasma channels. Carbon nanostructures have potential advantages over crystals for channeling acceleration such as wider channels (weaker de-channeling), broader beams (using nanotube ropes), wider acceptance angles (< 0.1 rad), 3D beam control over greater lengths, and in particular excellent thermal and mechanical strength, which are ideally fit to channeling acceleration and cooling applications, plus beam extraction, steering, and collimation. CNTs are comprised entirely of $sp2$ bonds, which are extremely stable and thermally and mechanically stronger than crystals, steels, and even diamonds ($sp3$ bond). Nanotubes thus have a higher probability of surviving in an extremely intense channeling, radiation, and acceleration environment.

Plasma frequencies of CNTs are normally in the THz range, which need a beam bunch duration within an order of microns for wakefield generation in the linear regime, an electron bunch in the pico-second range, which is readily obtained from conventional RF photoemission technique, would need to be either compressed down to an order of a femto-second or split into microbunches. It is well known that injecting a driver beam with multiple microbunches in a plasma channel improves field gradient, transformer ratio, and energy efficiency of plasma wakefield acceleration. With the multiple microbunches, phase matching conditions between a plasma wave and a modulated beam can be selectively tuned in order to maximize wakefield, transformer ratio, or energy efficiency.

## SIMULATION ANALYSIS

The basic pattern of channeling acceleration has been analyzed by theoretical model and computer simulation with the nominal ASTA beam parameters: bunch-to-bunch space = 10 μm, beam energy = 50 MeV, charge density of plasma channel = $10^{25}$ m$^{-1}$, bunch length = 2 ~ 3μm. Figure 2 shows summarized acceleration gradient and energy gain graphs with respect to beam charges, obtained from the 1D linear wakefield theory and plasma accelerating simulator (VORPAL). The simulation data agree well with the theoretical graph in the linear regime. This result clearly verifies that micro-spaced electron bunches gain energy up to ~ 70 MeV (~ 200 pC) along a 100 μm long channel, corresponding to a ~ 0.7 TeV/m acceleration gradient. The analysis indicates that further increase of the beam charge excessively blows out the ambient plasma charges in the channel ((3) of Fig. 3: bottom). It strongly pushes plasma waves out of phase-synchronization with the beam wave, which rather decreases the energy gain. The analyzed parameters will be implemented into the design process of the experimental setup. This kind of simulation method will be continuously used to analyze/design the beam-driven accelerator with various channeling conditions.

## EXPERIEMNTAL LAYOUT

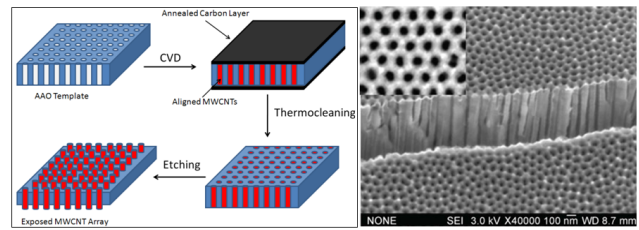

Figure 3: (a) Schematic diagram of an AAO template process (b) SEM image of an AAO film, inset is the enlarged SEM image.

Two kinds of channeling accelerator structures, crystals (silicon or diamond) and arrayed CNT bundles, will be manufactured and tested and compared in the first experiment. Crystal samples will be obtained by a similar fabrication technique used for silicon samples in channeling experiments in Tevatron of Fermilab: cleaved and cut silicon wafers [9]. The sample holder, used for the experiment, will also be re-used for crystal acceleration tests, but only with the straight section (no bending). A test device with a straight multi-wall CNT bundle will be prepared using an anodic aluminum oxide (AAO) template (Fig. 3(a)) [10]. We plan to test a 100 μm long, 200 nm wide CNT structure first, which may possibly be manufactured within 2 ~ 3 months. It is expected that these types of crystal channels induce plasmonic waves of $\lambda_p \leq \sim 10$ μm.

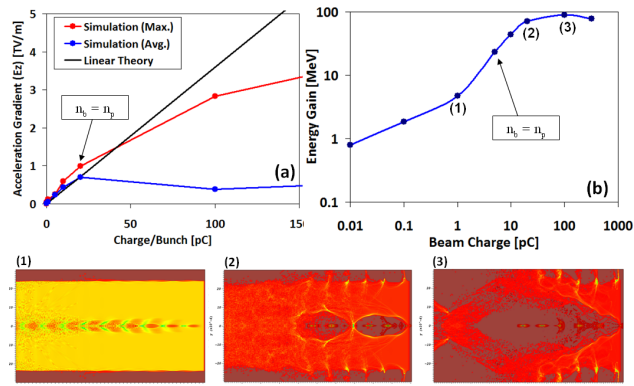

Figure 2: Channeling acceleration analysis, obtained from the linear wakefield theory and computer simulations: top - (a) acceleration gradients and (b) energy gain versus beam charge and bottom - spatial charge distributions of plasma and beam (1) 10 pC (2) 200 pC (3) 1 nC.

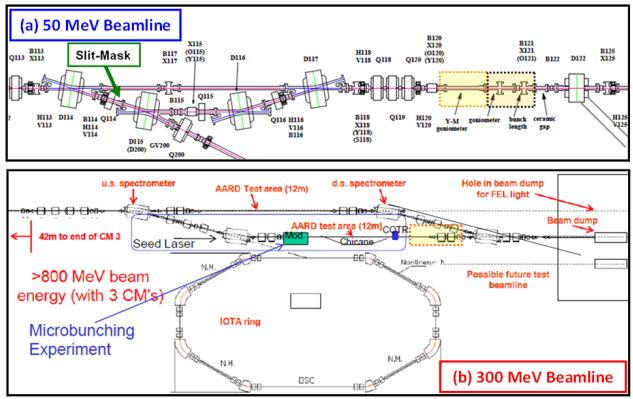

Figure 4: Prospective locations of experiment setup: 50 MeV beam line (top) and 300 MeV (bottom).

As we plan to conduct the 1st experiment at 50 MeV beamline of the Fermilab-ASTA, test equipment will be installed downstream of the capture cavity-2 (CC-2) and bunch compressor before the 1st cryomodule. Figure 4 shows the beam line drawing, describing the currently planned location of the proposed experiment. The boxed areas indicate prospective locations of the vacuum-compatible goniometer to be loaded with a test sample. While passing through the BC, the slit-mask placed in the BC will produce microbunch trains by imprinting the shadow of a periodic mask onto a bunch with a correlated energy spread. Once passing through a test device in the goniometer, beam energies will be measured by a spectrometer with a dipole magnet and yttrium aluminum garnet (YAG) screen positioned in the beamline toward a beam dump. A test with higher energy beams is also planned, so the experiment will be accommodated to the Development of a Compact Photo-injector with RF-Focusing Lens for Short Pulse Electron Source Application mainly consider the slit-mask technique, which is the simplest way to create multiple sub-ps micro-bunches [11], and an inverse free electron lasing technique. Currently, a laser-induced microbunching (LIM) scheme with the FEL undulator is proposed in the ASTA stewardship program, which is potentially employed to generate microbunches to the channeling acceleration test.

The proposed experimental scheme will need several positions to monitor bunch profiles and measure their spatiotemporal radiation patterns and energy spectra of accelerated beams. We measure the time pattern of a masked beam entering the dispersion-free region of the beam line using coherent transmission radiation (CTR) interferometry. The broadband transition radiation emitted by the electrons when entering a copper mirror placed after the dogleg, near the experimental region, is sent to a Martin-Puplett interferometer. After the energy-analyzed and collimated beam of electrons is transported through the bunch compressor into the experimental area, it is defocused by an asymmetrically split quadrupole triplet to give a low-divergence (nearly parallel) beam incident upon the crystal in its goniometer. A critical factor in performing channeling acceleration experiments is the divergence of the incident beam. Since the characteristic angle for the process is $1/\gamma$, an angular resolution at least an order of magnitude smaller is required in order to obtain data of sufficient precision to compare with the results of theoretical calculations; for $\gamma \sim 100$, a beam divergence larger than 1 mrad is inadequate. Moreover, the critical angle for channeling is a few mrad for electrons of a few tens of MeV, and varies as $\gamma^{-1/2}$; therefore, in order that a large fraction of the beam be channeled, a beam divergence $\leq 1$ mrad is required. The experimental arrangement is used for obtaining a very-low-divergence beam, which is used as well for measurements of the transmission of electrons through crystals.

## CONCLUSION

Despite the great potential of HEP colliders, excessively high driving energy and power requirements accompanied by the insufficient durability of crystal structures has removed channeling acceleration from primary consideration for high gradient (HG) accelerators. However, replacing crystals with nano-structures makes this possible by mitigating the power and energy requirements, with the advantage of improved physical strength. The fundamental mechanisms of plasmon excitation and photon-particle coupling of the nanotube can thus be studied at Fermilab-ASTA. Successful demonstration of all the simulations and experimental tests will open new opportunities for HG accelerator research efforts by merging nanotechnology and high energy physics. All of the new techniques and methods developed from heuristics will indeed be incorporated into existing technologies for current HEP collider R&D programs.

## ACKNOWLEDGEMENT

We thank Alex Lumpkin for his suggestion on micro-bunching technique. We thank Philippe R. G. Piot for helpful discussion and support for the proposal and Chris Prokop for sharing his ASTA simulation design parameters.